\documentclass[lettersize,journal]{IEEEtran}
\usepackage{amsmath,amsfonts}
\usepackage{algorithmic}
\usepackage{algorithm}
\usepackage{array}
\usepackage[caption=false,font=normalsize,labelfont=sf,textfont=sf]{subfig}
\usepackage{textcomp}
\usepackage{stfloats}
\usepackage{url}
\usepackage{verbatim}
\usepackage{graphicx}
\usepackage{cite}
\begin{document}


\title{Near-Field ISAC for THz Wireless Systems}

\author{Fan~Zhang, 
        Tianqi~Mao,~\IEEEmembership{Member,~IEEE}, 
        Mingkun~Li, 
        Meng~Hua,~\IEEEmembership{Member,~IEEE},
        Jinshu~Chen,\\
        Christos~Masouros,~\IEEEmembership{Fellow,~IEEE}, 
        Zhaocheng~Wang,~\IEEEmembership{Fellow,~IEEE}
\thanks{This work was supported in part by National Key R$\&$D Program of China under Grant 2021YFA0716600,
in part by National Natural Science Foundation of China under Grant 62401054 and Grant 62088101, and in part by Young Elite Scientists Sponsorship Program by CAST under Grant 2022QNRC001. \emph{(Corresponding author: Zhaocheng Wang and Tianqi Mao.)}} 

\thanks{F.~Zhang, M.~Li, J~Chen, and Z.~Wang are with Department of Electronic Engineering, Tsinghua University, Beijing 100084, China (e-mails: zf22@mails.tsinghua.edu.cn, lmk23@mails.tsinghua.edu.cn, chenjs@tsinghua.edu.cn, zcwang@tsinghua.edu.cn).} 
\thanks{T. Mao is with State Key Laboratory of CNS/ATM, Beijing Institute of Technology, Beijing 100081, China, and is also with Beijing Institute of Technology (Zhuhai), Zhuhai 519088, China (e-mail: maotq@bit.edu.cn).}

\thanks{M. Hua is with Imperial College London, SW7 2AZ London, U.K. (e-mail: m.hua@imperial.ac.uk).}

\thanks{C.~Masouros is with Department Electronics \& Electrical Engineering, University College London, London WC1E 7JE, UK (e-mail: c.masouros@ucl.ac.uk).}
}



\maketitle

\begin{abstract}
Sixth-generation (6G) wireless networks are expected not only to provide high-speed connectivity but also to support reliable sensing capabilities, giving rise to
the integrated sensing and communication (ISAC) paradigm.  
To enable higher data rates and more accurate sensing, terahertz (THz) systems empowered by extremely large multiple-input-multiple-output (XL-MIMO) technology are envisioned as key enablers for future ISAC systems. 
Owing to the substantial increase in both effective array aperture and carrier frequency, a considerable portion of future ISAC applications is anticipated to fall within the near-field coverage region, instead of the conventional far-field.
However, most existing ISAC techniques are designed under the far-field planar wave assumption, struggling to accommodate the unique characteristics of THz near-field propagation.
To motivate future research into near-field ISAC research, we systematically investigate the characteristics of THz near-field propagation and explore its potential to facilitate ISAC systems.
Specifically, we analyze three fundamental characteristics of THz near-field propagation and review state-of-the-art techniques that exploit these features to boost both communication and sensing performance.
To further harness the angular-range coupling effect, we zoom into a particularly interesting approach to near-field sensing based on wavenumber domain. Besides, to exploit the beam squint effect, an ISAC resource allocation framework is introduced to support integrated multi-angle sensing and multi-user communication.
Finally, we outline promising directions for future research in this emerging area.
\end{abstract}


\section{Introduction}
Driven by a wide range of emerging applications, such as virtual reality, smart cities, and the low-altitude economy, there is an increasing demand for wireless networks that not only ensure seamless connectivity but also provide high-resolution sensing capabilities \cite{bg1}. To meet these demands using existing network infrastructure, integrated sensing and communication (ISAC) has emerged as a promising paradigm for sixth-generation (6G) networks \cite{bg2}. By combining sensing and communication functionalities within a unified hardware framework, ISAC offers an innovative approach to enhancing spectrum efficiency while simultaneously reducing hardware complexity and cost \cite{Near-field1}.


With the explosive growth in mobile devices and data traffic, the terahertz (THz) spectrum is being actively explored to deliver substantial capacity gains.
Meanwhile, the ultra-high carrier frequency and broad bandwidth offered by the THz band enable fine-grained resolution in both velocity and range estimation. As a result, THz-enabled ISAC systems have garnered considerable attention from both industry and academia \cite{BS0, BS1}. To overcome the severe path loss inherent at THz frequencies and to achieve ultra-high spatial resolution, such THz-enabled systems are typically equipped with extremely large multiple-input-multiple-output (XL-MIMO) technology capable of forming highly directional beams, benefiting both communication data rates and angle estimation for sensing \cite{bg3}.


By employing XL-MIMO empowered THz systems in ISAC applications, the substantial increase in both array aperture and carrier frequency leads to an extended Rayleigh distance, which commonly defines the boundary between the near-field and far-field regions \cite{Near-field2}.
In the near-field region, the conventional planar wave assumption fails to accurately model spherical wave propagation, thereby giving rise to new characteristics in THz signal behavior \cite{Near-field3}. 
In this paper, we focus on three key features of THz near-field channels and signal propagation, namely the angular-range coupling, the near-field beam squint, and the flexible near-field beam controllability, respectively.
Conventional ISAC techniques based on the planar wave assumption, such as channel estimation, precoding, and range-angle sensing, exhibit performance limitations in THz near-field scenarios, struggling to approach optimal communication and sensing performance.

Although the unique characteristics of THz near-field propagation pose challenges to existing ISAC techniques, they also open up new opportunities for enhancing dual-function performance through innovative approaches \cite{BS2}. Against this backdrop, we systematically explore the effects of THz near-field propagation and investigate how these phenomena can be harnessed in ISAC systems. Furthermore, we invoke several interesting methods tailored for next-generation 6G ISAC systems operating in the near-field region. The main contributions of this paper are summarized as follows:
\begin{itemize} 
\item The fundamental principles of three key characteristics of THz near-field propagation are analyzed, and a comprehensive review of state-of-the-art technologies leveraging these features to enhance both communication and sensing performance is provided.
\item 
We highlight a near-field sensing technique based on the wavenumber domain, enabling accurate range estimation without requiring precise timing synchronization between the transmitter and receiver.
\item 
We explore a resource allocation framework for ISAC that leverages near-field beam squint to enable both multi-directional sensing and efficient communication.
\item 
Finally, we present an in-depth discussion of open research challenges and future directions for THz near-field ISAC design.
\end{itemize}

\section{Key Features of THz Near-Field Propagation}
\begin{figure*}[!t]
\centering
\includegraphics[width=0.85\linewidth]{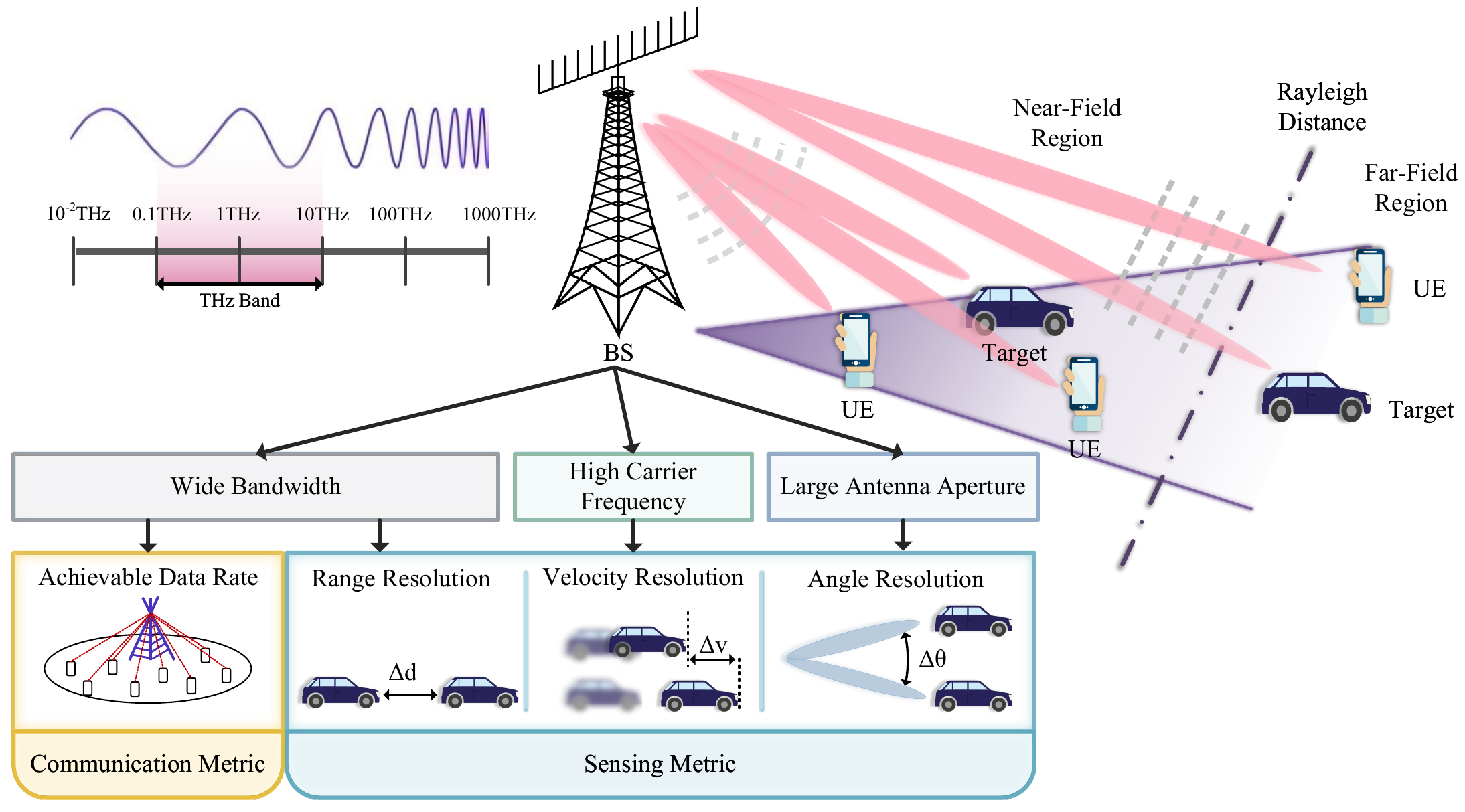}
\caption{Features and advantages of THz ISAC systems.}
\label{fig_1}
\end{figure*}
A typical XL-MIMO empowered THz ISAC system is illustrated in Fig.~\ref{fig_1}, where a base station (BS) forms directional beams for downlink data transmission to users (UEs), while simultaneously sending probing signals for sensing tasks.
Characterized by wide bandwidth, ultra-high carrier frequency, and large antenna aperture, such systems offer a variety of intrinsic advantages.
From the communication standpoint, the expanded bandwidth significantly increases the achievable data rate according to Shannon’s capacity formula. 
From the sensing perspective, the large bandwidth improves range resolution by refining delay-domain accuracy, while the ultra-high carrier frequency boosts velocity estimation accuracy by amplifying Doppler shifts. 
Additionally, the deployment of large-scale antenna arrays facilitates accurate angle estimation, owing to their superior spatial resolution.

However, these advantages cannot be fully realized by directly adopting existing ISAC techniques developed for the far-field scenarios, as THz systems empowered by XL-MIMO exhibit unique characteristics that require tailored system designs. 
One of the most critical distinctions lies in the extended Rayleigh distance, which causes a considerable portion of ISAC UEs and targets to reside in the near-field region. In this regime, the planar wave approximation becomes invalid, making it necessary to reevaluate propagation models and system strategies. This calls for a comprehensive investigation into the characteristics of THz near-field channels and their implications for ISAC design.
In this section, we identify and analyze three key features of THz near-field propagation, which are the angular-range coupling, the near-field beam squint, and the flexible near-field beam controllability, respectively. 
For each feature, we elaborate on its physical principles and discuss the limitations of existing technologies in addressing these phenomena.

\subsection{Angular-Range Coupling}
Consider an ISAC system operating at carrier frequency $f_c$ with a uniform linear array (ULA) consisting of $N$ antennas. Let $\mathbf{a}=[a(0),a(1),\cdots,a(N-1)]^{T}$ denote the antenna array response vector, where $a(n)$ denotes the response coefficient on the $n$-th antenna written as
\begin{align}
    a(n) = e^{-\textsf{j}{2\pi}{f_c}\sqrt{\tau^2+t_n^2-2\tau t_n \cos\theta}}.
    \label{array-response}
\end{align}
Here, $\tau$ and $\theta$ denote the delay and angle of arrival (AoA) of the UE/target relative to the ULA center, respectively. $t_n$ is the propagation time from the $n$-th antenna to the array center, given by $t_n = \frac{(n-1-(N-1)/2)d}{c}$, where $d$ is the antenna spacing and $c$ is the light speed.
When the UE/target is in the far-field region, i.e., $\tau \gg t_n$, $a(n)$ can be approximated as
\begin{align}
    a(n) = e^{-\textsf{j}{2\pi}{f_c}( \tau -t_n \cos\theta}),
    \label{array-response2}
\end{align}
which corresponds to the conventional far-field planar wave model.
A key distinction between (\ref{array-response}) and (\ref{array-response2}) lies in the structure of the exponential term. In (\ref{array-response}), the delay $\tau$ and antenna position $t_n$ are intertwined in a multiplicative cross-product, whereas in (\ref{array-response2}), $\tau$ and $t_n$ appear as independent additive components.
This difference becomes evident when transforming the response into the angular domain via a discrete Fourier transform (DFT). In the far-field model as (\ref{array-response2}), the angular representation is largely invariant with respect to the delay $\tau$.
In contrast, for near-field array response represented by (\ref{array-response}), its angular-domain representation is dependent of $\tau$. Figure~\ref{fig_2}(a) illustrates how the angular representation of the near-field channel changes with the BS–UE distance. As the range increases, noticeable variations in the angular-domain representation can be observed, which gives rise to the phenomenon referred to as the angle–range coupling effect.

Due to this coupling, the angular domain representation of near-field channel no longer exhibits favorable sparsity.
As a result, existing angle-domain-based channel estimation techniques suffer from main-lobe diffusion, as illustrated in Fig.~\ref{fig_2}(b), which impairs their ability to accurately capture the power of dominant paths and leads to increased estimation errors. 
Furthermore, DFT-based precoding schemes suffer from energy leakage due to the mismatch between codebooks and channel structures, resulting in degraded performance \cite{Near-field2}.

\subsection{Near-Field Beam Squint}
Due to the large bandwidth available in THz bands, the array response at each frequency point can not be approximated by that of the center frequency. Consequently, the beamforming patterns vary across different frequency points, leading to the phenomenon known as beam squint \cite{BS0}.

\begin{figure*}[!t]
\centering
\includegraphics[width=0.9\linewidth]{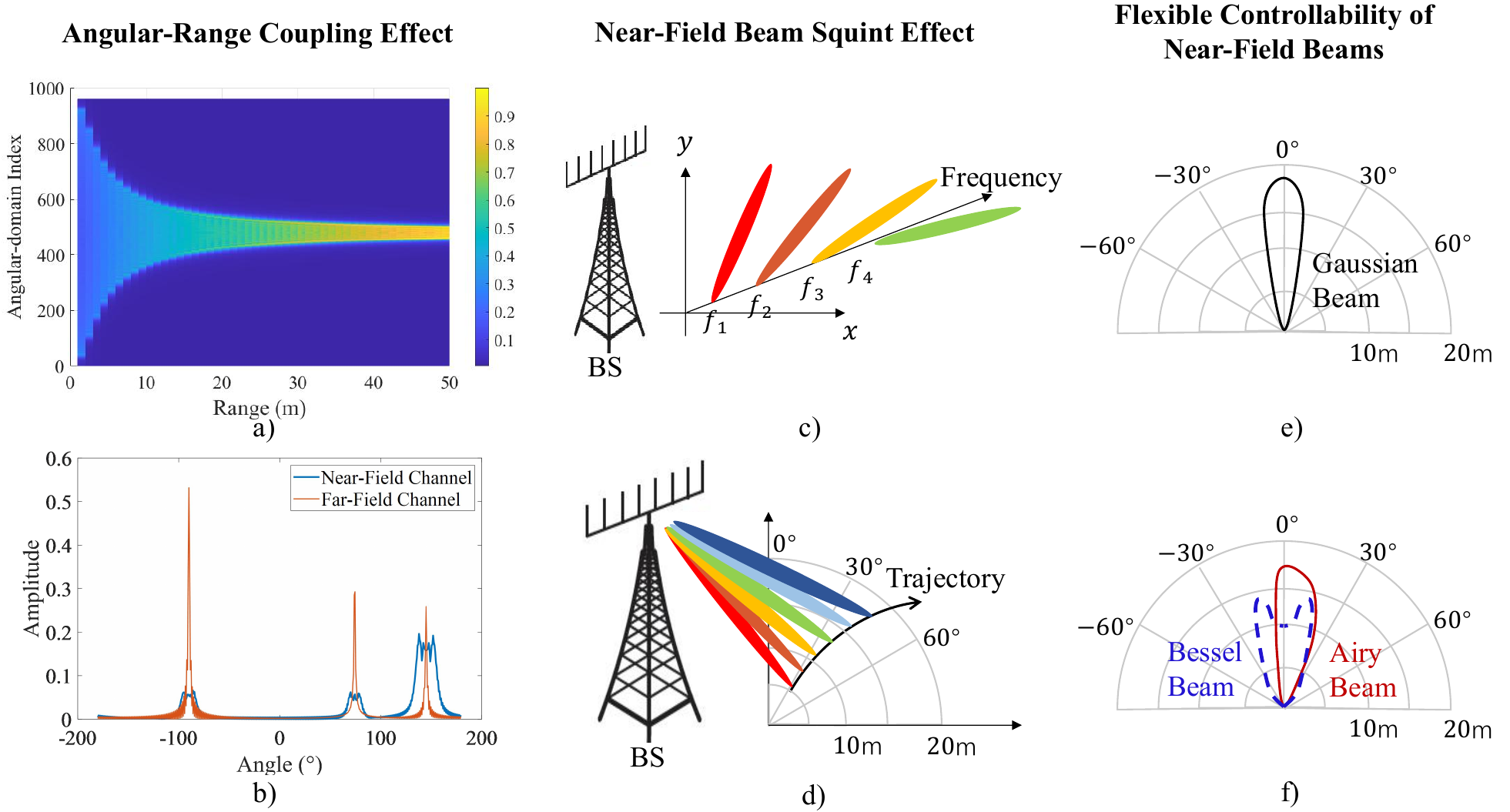}
\caption{Three key features of THz near-field propagation.}
\label{fig_2}
\end{figure*}

Consider orthogonal frequency division multiplexing (OFDM) as an example, which is a widely adopted multi-subcarrier waveform.
Suppose there are a total of $M+1$ subcarriers and the center frequency is $f_c$. The frequency of the $m$-th subcarrier is given by $f_c + (m-\frac{M}{2})\Delta f$, where $\Delta f$ denotes the subcarrier spacing.
The array response at the $n$-th antenna for the $m$-th subcarrier denoted as $a_m(n)$ can be expressed as 
\begin{align}
    a_m(n) = e^{-\textsf{j}{2\pi}{(f_c + (m-\frac{M}{2})\Delta f)}\sqrt{\tau^2+t_n^2-2\tau t_n \cos\theta}}.
    \label{array-response3}
\end{align}
When employing the DFT codebook for beamforming in the far-field region, each subcarrier's beam points toward a slightly different angle, as presented in Fig.~\ref{fig_2}(c), making the energy deviate from the desired UE/target position. The consequent loss in beamforming gain results in severe degradation of communication capacity and reduced accuracy in target detection.

In the near-field region, the phenomenon of beam squint becomes even more intricate. Beams associated with different subcarriers diverge not only in angle but also in range, as shown in Fig.~\ref{fig_2}(d). 
To illustrate this distance-dependent beam squint effect, we consider a commonly used polar-domain codebook, which is designed to form a beam focused at a specific point \((r, \theta)\) in the near field, where \(r\) denotes the distance from the antenna array and \(\theta\) represents the AoA. 
Due to the beam squint effect, when this codebook is applied to a wideband system, only the beam generated at the center frequency accurately aligns with \((r, \theta)\), while the beams associated with other subcarriers increasingly deviate in both angle and range.
For instance, consider a user or target located at \((60^\circ, 10\,\mathrm{m})\), the near-field beam squint effect can lead to an approximate \(7^\circ\) angular deviation and \(6\,\mathrm{m}\) range deviation when operating at a carrier frequency of \(300\,\mathrm{GHz}\) with a bandwidth of \(30\,\mathrm{GHz}\) \cite{BS1}. 

The focal points of beams formed by different subcarriers trace a continuous spatial curve, with the starting point and ending point corresponding to the focal points of the lowest and highest frequency subcarriers, respectively. We refer to this curve as the \textit{near-field beam squint trajectory}.
Most existing studies regard the near-field beam squint effect as a detrimental phenomenon and have proposed various compensation techniques for ISAC system design, such as subcarrier-dependent (SD) based phase shift networks and true time delay (TD) techniques. 
Nevertheless, the near-field beam squint effect also offers a unique opportunity for multi-target sensing across multiple directions. By precisely controlling the beam squint trajectory, it is possible to simultaneously sense multiple potential target locations without the need for frequent beam switching. This opportunity will be elaborated in detail in Section~\uppercase\expandafter{\romannumeral3}.

\subsection{Flexible Near-Field Beam Controllability}
According to the Huygens–Fresnel principle, the shape of a radiated beam is fundamentally determined by the spatial distribution of the electric field’s phase and amplitude across the radiating aperture \cite{Curving_THz1}. In the far-field region, the receiver experiences only negligible phase differences among the electromagnetic waves emitted from different elements of the antenna array due to the long propagation distance. As a result, the superposition of these wavefronts forms a well-collimated Gaussian beam directed along a single direction, as illustrated in Fig.~\ref{fig_2}(d).

In contrast, in the near-field region, where the observation plane is relatively close to the antenna aperture, the individual phase contributions of each array element become more pronounced. This spatial phase sensitivity enables a new degree of freedom. By precisely manipulating the phase profile at the source, it becomes possible to synthesize a wide variety of beam shapes and patterns. This property opens up new opportunities for precise beam shaping and spatially flexible communication and sensing in near-field systems. Particular opportunities arise with spot beams allowing extreme beam focusing in both range and angle and curved beams allowing NLOS propagation around objects.
Meanwhile, thanks to the extremely short wavelengths of THz signals, the radiating elements can be made very compact, enabling even moderately sized antenna panels to accommodate a large number of elements. This high element density enables fine-grained control over the phase across the aperture, enabling the generation of phase profiles that closely approximate continuous fields.
Such high-resolution beam control significantly enhances the flexibility of beamforming in the near-field region, enabling the realization of various non-Gaussian beam patterns that are unattainable in conventional far-field systems.

As shown in Fig.~\ref{fig_2}(e), Bessel beams and airy beams are typical non-Gaussian THz beams in the near field.
The Bessel beam exhibits a pronounced power drop in a specific direction, which enables the majority of the energy to reach the UE even in the presence of small obstacles.
Meanwhile, the inherent energy asymmetry of the airy beam enables it to propagate along a curved trajectory. This property facilitates the circumvention of large obstacles, thereby mitigating the signal blockages that are common in THz bands.
Owing to their distinctive propagation characteristics, these non-conventional beam shapes offer promising advantages for imaging and sensing tasks \cite{Curving_THz3}. A more detailed discussion of their applications will be presented in Section~\uppercase\expandafter{\romannumeral3}.


\section{Advances in ISAC Technologies Enabled by THz Near-Field Features}

In this section, we review state-of-the-art ISAC techniques that leverage the unique properties of THz near-field propagation.
Building upon these insights, we highlight two emerging methodologies aimed at enhancing ISAC performance. Firstly, to fully exploit the angular-range coupling effect, a wavenumber-domain sensing scheme is introduced to enable accurate range estimation without requiring precise delay synchronization. Secondly, by harnessing the beam squint effect, we present a resource allocation framework that supports integrated multi-angle sensing and multi-user communication, thereby improving the overall efficiency in near-field region.

\subsection{Exploiting Angular-Range Coupling for Enhanced ISAC Performance}
\textbf{Beam focusing:} 
Because of the angular-range coupling characteristics inherent in THz near-field propagation, the beam pattern becomes range-dependent. 
This spatial dependency allows the receiver to distinguish different signals in both angular and distance domains, thereby enhancing the signal multiplexing gain by increasing the spatial degrees of freedom (DoFs).
To fully exploit the spatial DoFs, beam focusing is proposed to concentrate the radiated energy at specific spatial locations \cite{Near-field1}.

Similarly to the conventional far-field beamforming, near-field beam focusing relies on digital and analog precoding techniques. 
Various precoding architectures have been proposed to achieve this, including fully digital architectures, hybrid phase-shifter-based precoders, and dynamic metasurface antenna (DMA) architectures. 
Fully-digital architectures provide the most flexible signal processing capabilities, achieving the highest data rate at the cost of high hardware overhead. 
Hybrid architectures, which combine both digital and analog signal processing, offer a practical compromise by reducing the number of required radio frequency (RF) chains and lowering implementation overhead.
DMA architectures employ radiating metamaterial elements to realize reconfigurable antennas of low cost and power consumption. 
Owing to their ability to support sub-wavelength element spacing, DMAs can pack a higher density of elements within a given aperture, thereby enabling highly focused beamforming in single-user scenarios \cite{Near-field2}.

By employing beam focusing in ISAC applications, mutual interference between sensing and communication can be effectively mitigated, even when the UE and the detected target are closely located in the angular domain. Furthermore, beam focusing enhances both communication capacity and sensing accuracy by increasing the signal-to-noise ratio (SNR) at the receiver. Finally, beam focusing has the ability to guarantee the security of signals from eavesdropping from other directions.


\textbf{Near-field target localization:}
In far-field scenarios, target localization is usually performed by the BS, which estimates the target's angle through beam sweeping and determines the range by calculating the delay between the transmitted signal and its echo. 
However, such monostatic ISAC architectures suffer from severe interference between the transmitter and receiver, as the reflected echo signal can be easily masked by the much stronger transmitted waveform. 
An alternative is multi-BS cooperative localization, which mitigate such interference but requires highly accurate synchronization among participating BSs, thereby introducing significant implementation challenges.


The angular–range coupling effect in THz near-field systems offers a novel approach to target localization. Specifically, once the channel state information is obtained, both the range and angle of the target can be inferred directly from the channel estimation results.
Consider a typical localization scenario where the target is a communication  UE located in the near-field region, equipped with a single omnidirectional antenna, and the BS is equipped with a large antenna array, such as a ULA or uniform planar array (UPA). The UE transmits a reference signal, which is received by the BS to extract the channel information. As demonstrated in \cite{Near-field3}, the range and angle can be jointly estimated from the received signal using the MUSIC algorithm. However, to accurately determine the target's three-dimensional (3D) position, the conventional MUSIC algorithm must be extended to a 3D search, resulting in significantly increased computational complexity.

\begin{figure}[!t]
\centering
\includegraphics[width=\linewidth]{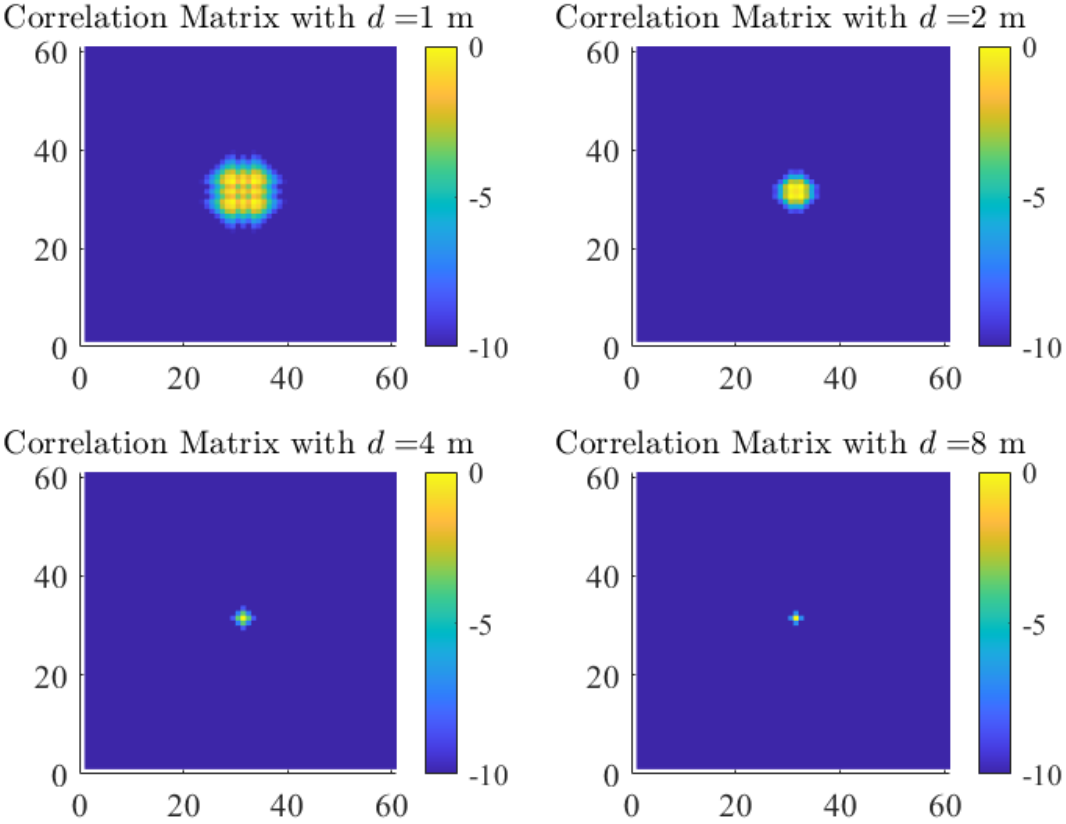}
\caption{The correlation matrix at the receiving end under different distances, where $d$ denotes the distance between BS and target.}
\label{fig_3}
\end{figure}


To address this issue, a low-complexity near-field ranging method can be conceived based on the wavenumber domain. By transforming the received signals into the wavenumber domain, both distance and angular information can be efficiently extracted from the resulting wavenumber domain pattern. The core concept of wavenumber domain transformation lies in approximating a spherical wavefront as a summation of planar waves. As illustrated in Fig.~\ref{fig_3}, the transformed signal yields a correlation matrix in which significant (non-zero) coefficients are concentrated within a circular region. The position of this circle indicates the direction of target, while its radius corresponds to the range of target. It can be observed from Fig.~\ref{fig_3} that as the target moves farther away, the radius of the circle decreases. This one-to-one mapping between target distance and radius enables accurate and efficient localization.

This approach enables target localization leveraging existing communication reference signals without requiring dedicated sensing waveforms. Meanwhile, it mitigates the need for precise synchronization and delay estimation. Overall, it introduces a promising new paradigm for the development of low-complexity and high-accuracy localization in future near-field ISAC systems.


\subsection{Beam Squint Assisted ISAC Systems}

\textbf{Integrated multi-angle sensing and multi-user communication:}
The near-field beam squint effect naturally provides directional frequency multiplexing for sensing and communications. As beams generated by different subcarriers are directed toward distinct spatial locations, it could simultaneously serve multiple UEs and multi-angle sensing \cite{BS3}.


The key to effectively leveraging the beam squint effect in ISAC applications lies in the precise control of beam directions across different subcarriers, enabling them to accurately point toward distinct UEs or targets \cite{BS4}. A widely adopted solution is the delay-phase array architecture, as illustrated in Fig.~4.  
In this architecture, the TD element introduces a frequency-dependent phase shift that varies linearly across subcarriers, while the phase shifter (PS) provides a constant phase offset across the entire bandwidth. By cascading TD and PS elements for each antenna path, the overall beam squint pattern can be flexibly adjusted to match the desired beamforming profile. For a given beamforming profile, the corresponding TD coefficient and phase offset can be determined by formulating a beam error minimization problem and solving it using the least squares method \cite{BS2}.

\begin{figure*}[!t]
\centering
\includegraphics[width=1\linewidth]{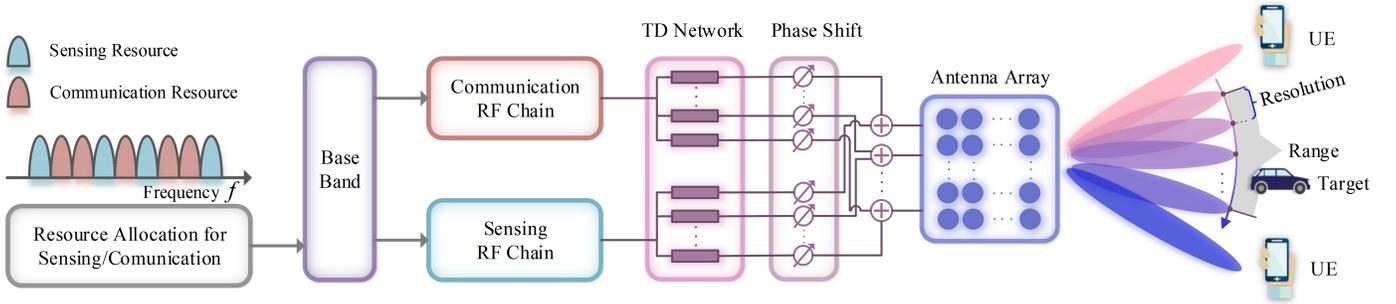}
\caption{Beam squint assisted ISAC resource allocation and hardware architecture.}
\label{fig_4}
\end{figure*}

Building upon the aforementioned architecture, a beam squint assisted framework can be conceived to integrate target sensing and multi-user communication, as illustrated in Fig.~\ref{fig_4}. 
Firstly, frequency resources and the power budget are allocated based on the requirements of communications and sensing systems, dividing the subcarriers into communication and sensing subcarriers. Specifically, the subcarrier assignment and power allocation can be formulated as an optimization problem to maximize the sum data rate of all the communication users while ensuring the required sensing range and accuracy.
The communication subcarriers can be further partitioned among users based on their individual demands. 
After modulation with communication symbols, the communication subcarriers are processed by the communication RF chain, TD network, and PS, directing each subcarrier toward its intended user.
Meanwhile, the sensing subcarriers are routed through a dedicated sensing RF chain, TD network, and PS to form beams that align with the predicted trajectory of potential target locations. In order to enhance sensing accuracy, prior target location information is utilized for trajectory prediction via Kalman filtering, and the delay and phase coefficients are dynamically adjusted based on the predicted results.
Although it is feasible to share a common RF chain between   communication and sensing subsystems to reduce hardware cost, this may compromise beamforming flexibility and introduce mismatches between the desired and actual beam patterns.
 
\begin{figure}[!t]
\centering
\includegraphics[width=.9\linewidth]{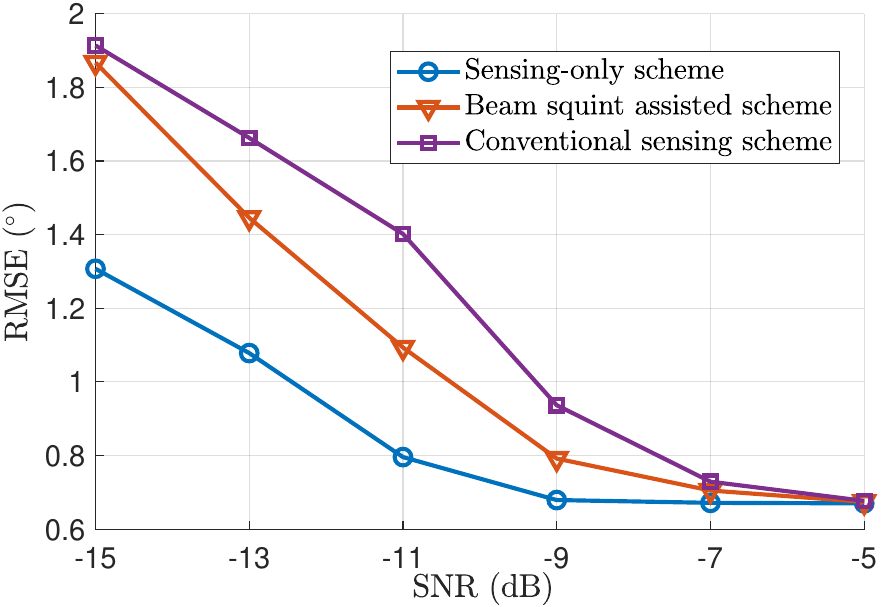}
\caption{RMSE comparison between the near-field beam squint assisted scheme and existing schemes with respect to SNR.}
\vspace{-3mm}
\label{fig_5}
\end{figure}

Figure~5 compares the angular estimation performance of the beam squint assisted scheme against existing benchmarks under different SNR levels, where the target is assumed to locate along an arc segment between \((60^\circ, 20\,\mathrm{m})\) and \((80^\circ, 20\,\mathrm{m})\).
The root mean square error (RMSE) is utilized as the performance metric.
In Fig.~\ref{fig_5}, a lower bound of RMSE is provided by the sensing-only scheme, where all available subcarriers are dedicated for radar detection. For the conventional scheme, the BS emits the probing signals to each direction sequentially, which requires frequent beam
sweeping to achieve fine-grained angle estimation in the entire sensing range.
In contrast, by exploiting the beam squint effect, the ISAC system can simultaneously sense multiple positions in one time slot.
As it can be seen in Fig.~5, the beam squint assisted scheme outperforms its conventional counterpart in term of RMSE with reduced beam switch cost, which validates its superiority.
Meanwhile, simulation results indicate that the beam squint assisted scheme is capable of achieving up to $95\%$ of the optimal communication data rate, while its sensing performance asymptotically approaches that of the sensing-only scheme as the SNR increases.

\subsection{Leveraging Flexible Near-Field Beams in ISAC}
\textbf{Beam manipulation for obstacle avoidance:}
To overcome the severe free-space path loss inherent in THz bands, highly directional beams are employed to enable reliable communication in 6G networks. However, such highly directional propagation is highly susceptible to blockage caused by user mobility and environmental dynamics, posing significant challenges for the deployment of THz systems in indoor wireless communication scenarios.
Fortunately, when both users and obstacles are located within the near-field region of the BS, advanced beam manipulation techniques can be leveraged to dynamically steer beams around obstructions, thereby facilitating reliable connectivity through proactive obstacle avoidance. 

The Bessel beam presents a promising solution for circumventing relatively small obstacles, whose amplitude distribution follows a Bessel function. A key characteristic of THz Bessel beams is their self-healing property. When partially obstructed, the beam can reconstruct its original pattern beyond the obstruction, which highlights the superiority of Bessel beams over conventional Gaussian beams in mitigating the impact of localized blockages in THz communication and sensing systems.
While Bessel beams perform well against minor obstructions, their effectiveness diminishes in the presence of larger obstacles. 
In such scenarios, airy beams demonstrate superior performance due to their distinctive property of self-acceleration. By imposing a cubic phase profile added with an auxiliary quadratic phase component at the radiating aperture, the resultant airy beam naturally follows a curved parabolic trajectory that enables it to bypass obstacles. This curved propagation arises from the constructive interference of caustics, i.e., collections of light rays that tangentially intersect the intended trajectory and guide the beam along its bending path \cite{Curving_THz1}.
Beyond obstacle avoidance, the unique propagation characteristics of Bessel and airy beams also play a crucial role in enhancing physical layer security, as they enable the transmitted signals from the BS to effectively circumvent potential eavesdroppers and reach the intended receiver \cite{Curving_THz2}.


\textbf{Near-field imaging and environment sensing:}
The high degree of beam controllability enabled by near-field THz technologies lays a solid foundation for advanced THz imaging and deep environmental sensing. 
By superimposing multiple complex electric fields corresponding to separate focal points, it is possible to generate multi-focal beams that simultaneously focus energy on several regions of interest across the surface of target.
This multi-focus capability significantly accelerates the scanning process and improves spatial resolution in THz imaging, thereby facilitating accurate surface profile reconstruction.
Additionally, the extended focal depth beam can concentrate energy along a focal line spanning the target’s depth dimension.
Such beams enable the extraction of two-dimensional characteristics, including target thickness and internal structural features \cite{Curving_THz4}.
These capabilities hold great promise for emerging applications such as high-precision 3D mapping and environment reconstruction.

\section{Open Challenges and Future Directions}
In this section, we outline several open challenges and corresponding research opportunities in THz near-field ISAC systems, from the perspective of waveform design, hardware implementation, and unified near/far-field ISAC schemes.

\textbf{Waveform design:}
In conventional far-field ISAC systems where range estimation relies on the echo delay derivation, the auto-correlation property of the waveform serves as a critical metric.
As a result, a long-standing challenge in ISAC waveform design is maintaining desirable auto-correlation characteristics in the presence of randomized communication symbols.
In contrast, near-field ISAC approaches can leverage channel state information for range estimation. The core challenge of ISAC waveform design may deviate from the conflict between auto-correlation property and randomness, and focus on fully harnessing the angular-range coupling effect inherent in near-field propagation.
As a result, the existing waveform designs may prove inadequate for meeting the unique demands of near-field sensing. Therefore, new theoretical frameworks and design methodologies are required for the next-generation ISAC waveform.

\textbf{Unified far-field and near-field ISAC:}
Due to the distinct propagation characteristics, near-field and far-field scenarios require different ISAC architectures.
Specifically, in the far-field region, angular-domain beamforming techniques are widely used for both data transmission and target sensing. In contrast, the near-field region benefits from beam-focusing and other flexible beam pattern designs enabled by precise phase control.
However, angular-domain beamforming suffers from power dispersion in near-field channels, while many near-field beam patterns are unattainable in the far-field. 
These limitations highlight the need for unified ISAC strategies capable of supporting both regions simultaneously.
Future research should investigate hybrid-compatible solutions, including unified codebook designs, advanced precoding techniques, and sensing algorithms that are adaptable to both near-field and far-field environments, enabling seamless communication and accurate sensing across both regions.

\textbf{Hardware implementation:}
To fully leverage the unique features of THz near-field channels, beam patterns must be precisely manipulated, which poses challenges for hardware implementation. 
For instance, achieving fine-grained control over range and angular resolution in beam squint–based systems requires TD elements that provide consistent delay responses across wide frequency bands. This requirement imposes substantial complexity on hardware implementation. 
Besides, the beam manipulation technology relies on phase control of electric field across the radiating aperture, which is typically realized through passive lenses.
However, each customized lens supports only one specific beam manipulation pattern and lacks reconfigurability.
To address this issue, reconfigurable intelligent surfaces have emerged as a promising alternative, which allows for the realization of various beam manipulation goals, but at the cost of high complexity.
The trade-off between reconfigurability and implementation complexity remains a critical research problem requiring further investigation.

\section{Conclusion}
In this article, we investigated the unique features and potential of XL-MIMO empowered THz ISAC systems.
We began by analyzing three key characteristics of THz near-field propagation. Building on these insights, we provided a comprehensive survey of existing technologies that leverage these features to improve the performance of dual-function systems. In addition, we zoomed into an interesting near-field sensing approach based on the wavenumber domain, along with a beam squint assisted ISAC resource allocation framework. Simulation results demonstrate that the framework achieves high-resolution angle estimation with low overhead while simultaneously supporting high-rate data transmission. Finally, we outlined several open challenges in waveform design, hardware implementation, and unified near/far-field ISAC strategies, offering promising directions for future research.

\vfill


\begin{thebibliography}{1}
\bibliographystyle{IEEEtran}

\bibitem{bg1}
K.~Meng, C.~Masouros, K.~-K.~Wong, A.~P.~Petropulu, and L.~Hanzo, ``Integrated sensing and communication meets smart propagation engineering: Opportunities and challenges," \emph{IEEE Network}, vol.~39, no.~2, pp.~278--285, Mar.~2025.
\bibitem{bg2}
F.~Zhang, \emph{et~al.}, ``Cross-domain dual-functional OFDM waveform design for accurate sensing/positioning,'' \emph{IEEE J. Sel. Areas Commun.}, vol.~42, no.~9, pp.~2259--2274, Sep.~2024.

\bibitem{Near-field1}
J.~Zhao, S.~Xue, K.~Cai, X.~Mu, Y.~Liu, and Y.~Zhu, ``Near-field integrated sensing and communications for secure UAV networks", \emph{arXiv preprint}, arXiv: 2502.01003, Feb.~2025. 


\bibitem{BS0}
Z.~Liu, C.~Yang, and M.~Peng, "Integrated sensing and communications in terahertz systems: A theoretical perspective," \emph{IEEE Network}, vol.~38, no.~3, pp.~194--201, May~2024.

\bibitem{BS1}
H.~Luo, F.~Gao, W.~Yuan, and S.~Zhang, ``Beam squint assisted user localization in near-field integrated sensing and communications systems,'' \emph{IEEE Trans. Wireless Commun.}, vol.~23, no.~5, pp.~4504--4517, May~2024.

\bibitem{bg3}
Q.~Dai, \emph{et~al.}, ``A tutorial on MIMO-OFDM ISAC: From far-field to near-field,'' arXiv e-prints, arXiv:2504.19091, Apr.~2025.

\bibitem{Near-field2}
Z.~Wang, X.~Mu, and Y.~Liu, ``Rethinking integrated sensing and communication: When near field meets wideband,'' \emph{IEEE Commun. Mag.}, vol.~62, no.~9, pp.~44--50, Sep.~2024.


\bibitem{Near-field3}
Y.~D.~Huang, and M.~Barkat, ``Near-field multiple source localization by passive sensor array,'' \emph{IEEE Trans. Antennas Propag.}, vol.~39, no.~7, pp.~968-975, Jul.~1991.



\bibitem{BS2}
J.~Li, S.~Zhang, Z.~Li, J.~Ma, and O.~A.~Dobre, ``User sensing in RIS-aided wideband mmWave system with beam-squint and beam-split," \emph{IEEE Trans. Commun.}, vol.~73, no.~2, pp.~1304--1319, Feb.~2025.

\bibitem{Curving_THz1}
H.~Guerboukha, B.~Zhao, Z.~Fang, E.~Knightly, and D.~M.~Mittleman, ``Curving THz wireless data links around obstacles'', \emph{Commun. Eng.}, vol.~3, no.~1, Mar.~2024. 

\bibitem{Curving_THz3}
M.~Li, J.~M.~Jornet, D.~M.~Mittleman, and C.~Han, ``Beam manipulation for terahertz communications: A new quality productive force,'' arXiv e-prints, arXiv:2503.22158, Mar.~2025.



\bibitem{BS3}  
F.~Zhang, \emph{et~al.}, ``Multicarrier waveform design for mmWave/THz integrated sensing and communication,'' in \emph{Proc. IWCMC 2024} (Ayia Napa, Cyprus), May~26-31, 2024, pp.~501--506.

\bibitem{BS4}
Z.~Li, Z.~Gao, and T.~Li, ``Sensing user's channel and location with terahertz extra-large reconfigurable intelligent surface under hybrid-field beam squint effect," \emph{IEEE J. Sel. Topics Signal Process.}, vol.~17, no.~4, pp.~893--911, Jul.~2023.


 

\bibitem{Curving_THz2}
V.~Petrov, H.~Guerboukha, A.~Singh, and J.~M.~Jornet, ``Wavefront hopping for physical layer security in 6G and beyond near-field THz communications,'' \emph{IEEE Trans. Commun.}, vol.~73, no.~5, pp.~2996--3012, May~2025.



\bibitem{Curving_THz4}
B.~Chung, D.~Headland, and W.~Withayachumnankul, ``Terahertz imaging with 3D-printed risley-prism and telecentric objective,'' \emph{IEEE Trans. Terahertz Sci. Technol.}, vol.~14, no.~4, pp.~446--454, Jul.~2024.


\end{thebibliography}
\end{document}